\documentclass[%
 reprint,%
 aip, jap,
superscriptaddress,
floatfix
]{revtex4-1}

\usepackage{makeidx}

\usepackage{graphicx}
\usepackage{amsmath}
\usepackage{epstopdf}
\usepackage{textcomp}
\usepackage{epstopdf}

\usepackage{ulem} 
\usepackage{color}

\usepackage{natbib}
\usepackage{bm}%
\usepackage[colorlinks=true,linkcolor=blue]{hyperref}%
\expandafter\ifx\csname package@font\endcsname\relax\else
 \expandafter\expandafter
 \expandafter\usepackage
 \expandafter\expandafter
 \expandafter{\csname package@font\endcsname}%
\fi
\makeindex

\begin{document}

\title{\color{blue}Influence of the magnetic field on the plasmonic properties of transparent Ni anti-dot arrays.}

\author{Emil Melander}

\author{Erik \"{O}stman}

\author{Janine Keller}

\author{Jan Schmidt}

\author{Evangelos Th. Papaioannou}
\thanks{Author to whom correspondence should be addressed. Email: vangelis@physics.uu.se}

\author{Vassilios Kapaklis}

\author{Unnar B. Arnalds}

\affiliation{Department of Physics and Astronomy, Uppsala University,
Box 516, SE-751 20 Uppsala, Sweden}

\author{B. Caballero}\affiliation{IMM-Instituto de
Microelectr\'onica de Madrid (CNM-CSIC), Isaac Newton 8, PTM, Tres
Cantos, E-28760 Madrid, Spain} \affiliation{Departamento de
F\'{\i}sica Te\'orica de la Materia Condensada, Universidad
Aut\'onoma de Madrid, 28049 Madrid, Spain.}

\author{A. Garc\'{\i}a-Mart\'{\i}n}
\affiliation{IMM-Instituto de Microelectr\'onica de Madrid
(CNM-CSIC), Isaac Newton 8, PTM, Tres Cantos, E-28760 Madrid,
Spain}

\author{J. C. Cuevas}
\affiliation{Departamento de F\'{\i}sica Te\'orica de la Materia
Condensada, Universidad Aut\'onoma de Madrid, 28049 Madrid,
Spain.}

\author{Bj\"orgvin Hj\"{o}rvarsson}

\affiliation{Department of Physics and Astronomy, Uppsala University,
Box 516, SE-751 20 Uppsala, Sweden}

\date{\today}

\begin{abstract}
Extraordinary optical transmission is observed due to the excitation of surface plasmon polaritons (SPPs) in 2-Dimensional hexagonal anti-dot patterns of pure Ni thin films, grown on sapphire substrates. A strong enhancement of the polar Kerr rotation is recorded at the surface plasmon related transmission maximum. Angular resolved reflectivity measurements under an applied field, reveal an enhancement and a shift of the normalized reflectivity difference  upon reversal of the magnetic saturation (transverse magneto-optical Kerr effect-TMOKE). The change of the TMOKE signal  clearly shows the magnetic field modulation of the dispersion relation of SPPs launched in a 2D patterned ferromagnetic Ni film.
\end{abstract}



\maketitle

Magneto-plasmonics offer unique possibilities to manipulate light by the use of external magnetic fields.\cite{Temnov, Belotelov,
Ctistis2009, PhysRevB.81.054424} The prevailing choice of materials for
fabrication of magneto-plasmonic structures has been combined structures of noble and magnetic metals /
dielectrics, such as Au and Co / Iron garnet.\cite{Temnov,
martin-becerra:183114, 1367-2630-10-10-105012} 
The basic idea behind this choice is the combination of the large plasmon activity of noble metals with the magnetic functionality provided by the additional materials. 
Another reason for the use of noble metals is the excellent resistance to oxidation, which is required to obtain durable patterned thin films. 
Ni is an interesting candidate in this context as it forms a thin and self-passivating oxide layer (approximately 1nm).\cite{poulopoulos:202503, Saiki199333}  Furthermore, the magneto-optical activity of Ni-based nano-patterns can be enhanced by the presence of surface plasmon polaritons (SPPs).\cite{Papaioannou:11, doi:10.1021/nl2028443, Nogues2011,grunin:261908,torrado:193109} 

The magnetic field can provide the means for control of SPPs, as been predicted for noble metals, \cite{PhysRevB.77.205113} and has been explored experimentally in hybrid structures.\cite{Belotelov,martin-becerra:183114, 1367-2630-10-10-105012} Early studies on this effect were targeted towards semiconductor-based  SPPs \cite{Hartstein19741223} but not in metallic systems, where high magnetic fields are required.\cite{PhysRevB.76.153402} In pure magnetic materials the need for high fields is not present as the magneto-optical effects are sufficiently strong.

In this letter we discuss the influence of an external magnetic field on the SPPs for the case of a pure magnetic metal, such as Ni, patterned in two-dimensions (2D) on a transparent substrate. We examine to what extent the ferromagnetic Ni can be used as  host material for SPPs.
We show that the magnetic field induces a modulation of the dispersion of SPPs excitation in Ni.

\begin{figure}
 \includegraphics[width =0.9\columnwidth]{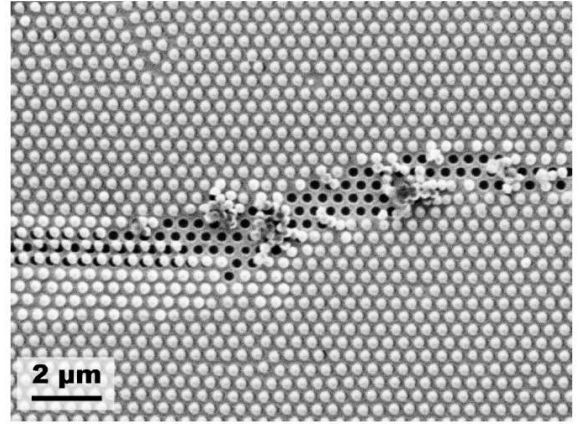}
    \caption{\label{fig_1} Scanning electron microscopy image of the sample surface after evaporation of Ni. The image depicts the stage of the lift off of the self-assembled polystyrene beads, which leaves behind an antidot patterned film of Ni. The resulting sample has an average pitch size of $a = 450 $ nm and hole diameter of $d = 300 $ nm.}
    \label{sem}
\end{figure}

A Ni anti-dot sample was prepared on a double side polished Al$_{2}$O$_{3}$ [11\=20] substrate. The patterning was accomplished by the use of self-organization of colloidal polystyrene beads as shadow masks.\cite{Papaioannou:11} A 30 nm thick Ni film was deposited on the masked sapphire substrate, using electron-beam evaporation. A snapshot of the procedure is illustrated in Fig.~\ref{fig_1}, where  both the shadow mask and the resulting holes are clearly seen.  This process resulted in a well defined Ni layer, decorated by holes of a diameter $d = 300$ nm, spaced on an hexagonal lattice of periodicity of $a = 450$ nm. The ratio of the radius to pitch size was determined to be  $\pi d^2/(2\sqrt{3} a^2) = 0.40$, leaving a total surface for the Ni film of 60 $\%$, with respect to the substrate area.

Magneto-optic spectra were recorded using a magneto-optic Kerr spectrometer operating in the polar configuration with an angle of
incidence of 4\textdegree\space and a maximum  applied  magnetic field of 1.7 T. The range in recorded wavelength is $250 - 1000$ nm. Angular dependent zero-order reflectivity curves
($\theta - 2\theta$) were obtained using a dedicated optical diffractometer (HUBER MC 9300), with a step resolution of 1/1000
degree and a laser wavelength of $\lambda$  = 660 nm \cite{Papaioannou:11}. A magnetic field up to 42 mT can be applied, using a quadrupole air coil magnet, which is mounted on the sample rotation stage of the goniometer.

\begin{figure}
 \includegraphics[width =1.0 \columnwidth]{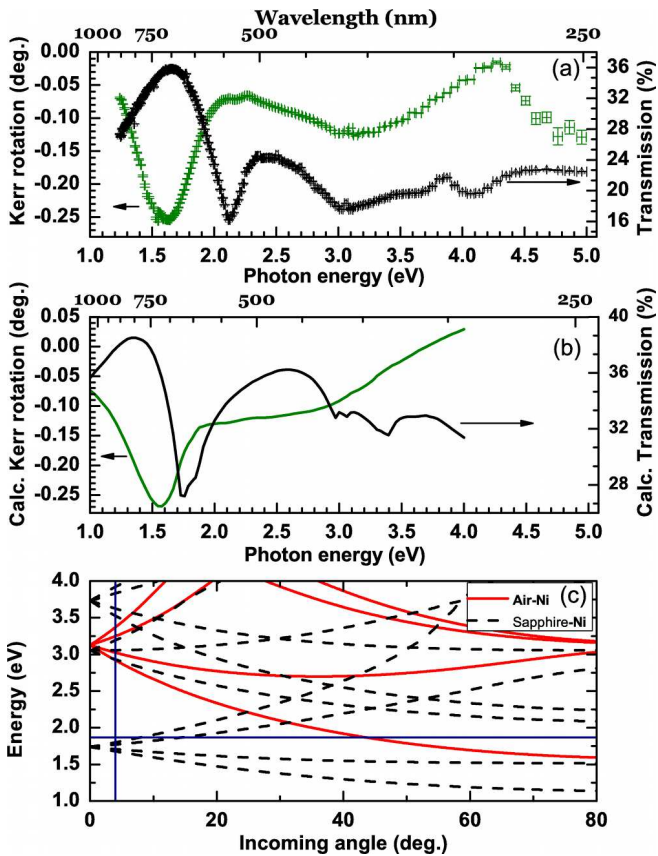}
\caption{\label{fig_2} (Color online) (a) Experimental Transmission and Polar Kerr
rotation spectra for the Ni antidot structure of
Fig.~\ref{fig_1}. Kerr values were obtained with the film at the
saturated state (B = 1.7 T). The transmission curve has been
corrected for the substrate. (b) Calculated Transmission and Polar Kerr
rotation spectra. (c) Calculated dispersion relation
of the Bragg plasmons both for the Ni-air interface and
Ni-sapphire interface. The vertical line is drawn at 4\textdegree \space  and the horizontal line at 660 nm.}
\end{figure}

Fig.~\ref{fig_2}(a) shows the experimental and (b) the calculated transmission and polar Kerr rotation
spectra for the sample. The calculations have been made using the scattering formalism described
in Ref.~\onlinecite{antonio05}, using the optical and magneto-optical elements of the dielectric tensor given in Ref.~\onlinecite{visnov93},
extended with those of Ref.~\onlinecite{mok2011} to include energies below 1.5 eV.

The experimental transmission curve in Fig. ~\ref{fig_2}(a) exhibit several
maxima. The calculated one in Fig. ~\ref{fig_2}(b)  shows a similar shape like in the experiment. 
If SPP modes are involved, then the energies at which the
maxima appear must be close to the energies corresponding to the
so called Bragg plasmons (SPPs modes coupled to the lattice
periodicity). The calculated dispersion relation of those Bragg plasmons is presented in Fig.~\ref{fig_2}(c) finding
a fair match for the transmission maxima.  As we move to higher frequencies the absorption plays more and more a significant role and that implies a shift and a widening of the resonance peaks. In addition as we can see from Fig. ~\ref{fig_2}(c), at higher frequencies, modes from the upper and lower interface begin to mix resulting in an overlapping of different resonances. Additional calculations (not shown) revealed that the coupling between the so-called low-index and the high-index plasmons at each interface is negligible.

The overlap of the
overall transmission maxima and the enhancement of the Kerr
rotation, further prove the existence of SPPs in the sample.
\cite{Ctistis2009, PhysRevB.81.054424, Papaioannou:11} The first
peak in transmission records an intensity of $\sim$ 36 \% (after
substrate corrections). Although it is expected the 
plasmon losses in Ni to be large, we have a clear
indication of a strong SPP resonance.  The propagation length of
SPPs in Ni  at 1.6 eV \cite{Palik} is calculated to be $840$ nm and decreases
strongly with increasing energy, resulting in low transmission
values as in Fig.~\ref{fig_2}(a). Transmission minima are related
to so-called Wood-Rayleigh anomalies.\cite{Ebbesen98} In a good metal the spectral location of the Wood-Rayleigh anomaly is close to the condition for SPPs excitation on a metal-dielectric interface. In our case, plasmon excitation occurs at slightly smaller energies (larger wavelengths) than the appearance of a diffracted beam. As a result maxima and minima are close to each other.

\begin{figure}
\includegraphics[width = \columnwidth]{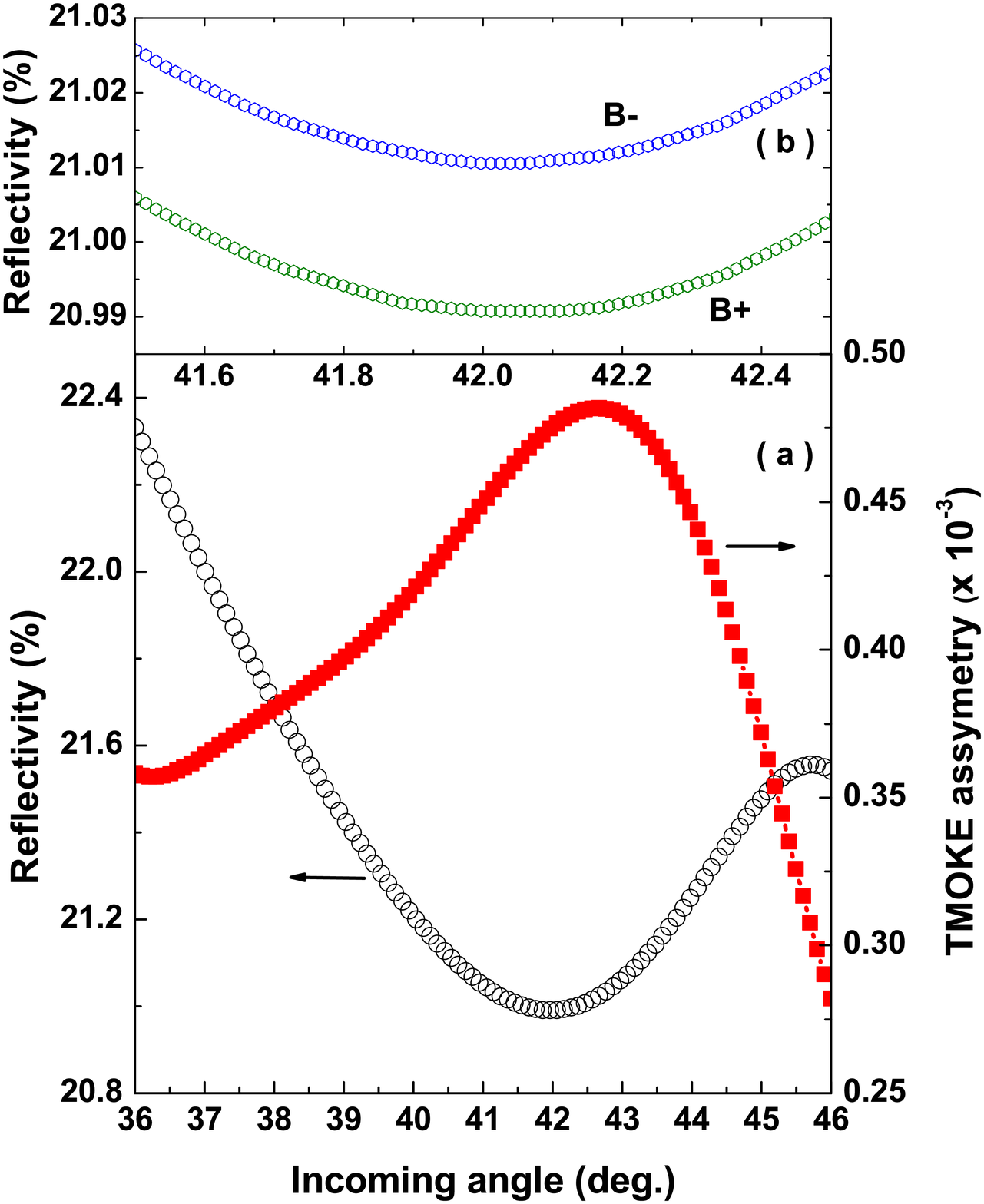}
\caption{\label{fig_3} (Color online) (a) Reflectivity
(black dot-line, left axis) and the TMOKE (red square-line, right axis). The
TMOKE signal presents a clear shift and enhancement. (b) Reflectivity in the
angular range of the minimum in the $\varGamma K$ direction
for the two directions of the magnetization.}
\end{figure}

Angular resolved reflectivity measurements are also a way to
explore the effect of SPPs in anti-dot structures.
\cite{Papaioannou:11} In Fig.~\ref{fig_3}(a)
(left y-axis) we present angular reflectivity
measurements at a wavelength of $\lambda = 660$ nm for p-polarized
light. This corresponds to the region in between the first transmission
maximum in Fig.~\ref{fig_2}(a) and the transmission minimum, or in
other words around a region with a large variation of
T($\lambda$). The data were obtained having the hexagonal pattern
aligned with one of its major symmetry axes ($\varGamma K$)
parallel to the scattering plane. The trough in reflectivity at
42\textdegree \space is a signature of surface plasmons
excitations. The minimum in reflectivity is close to the
theoretical value for the Bragg plasmon at that specific
frequency  ($\sim 1.85$ eV), as shown by the horizontal line in Fig.
\ref{fig_2}(c). The width of the reflection minimum is broader than
those for noble metals due to the higher absorption losses of Ni.
There is an additional broadening due to the ratio of the hole
depth (30 nm) to the hole diameter (300 nm). If the ratio is close
to unity the features are sharper than for lower ratios.
\cite{Ebbesen98}

In order to investigate any influence of the magnetic field on the
SPPs resonances, the transverse magneto-optic Kerr effect (TMOKE)
can be a very useful tool. The orientation of the
magnetic field with respect to the SPPs propagation direction and
the light polarization is of key importance\cite{Belotelov,
martin-becerra:183114} defining the magnitude of the magnetic
field - SPPs interaction.  TMOKE is defined as the magnetization
modulated intensity difference of the reflected light:

\begin{equation}
 \text{TMOKE} = \frac{R(M+)-R(M-)}{R(M+)+R(M-)}
\end{equation}

The applied magnetic field lies in the plane of the film and
perpendicular to the plane of incidence (and in our case also
perpendicular to the SPPs propagation along the $\varGamma K$
direction). In this configuration the magnetic field
influences only the optical reflectivity. The size of the applied
magnetic field was 20 mT, enough for the magnetic saturation of
the sample. Worth noticing, is that the magnetization of the Ni
nano-patterned film lies in the film plane (saturation field of 14
mT for the transverse case) while a high field of 0.45 T is needed
to saturate the sample out of plane.

The application of a transverse field induces a small change in the
position of the minimum  and the width of the resonant position of
the SPPs, and a variation of the intensity of the overall
reflectivity (see Fig.~\ref{fig_3}(b)). In the case of a ferromagnetic dielectric
film\cite{Belotelov:09,Belotelov} or Co layer
\cite{martin-becerra:183114} covered by a thin smooth/or
perforated noble metal  a transverse applied field shifts  the
plasmon related reflectivity (transmission) minimum (maximum) and
an enhanced TMOKE appears. A
similar behavior is observed here in the Ni film, having a
2D nano-pattern without the support of a noble metal. The TMOKE
signal presented as a red square-line in Fig.~\ref{fig_3}(a) appears
shifted with regards to the reflectivity minimum and enhanced
compared to the featureless TMOKE (not shown) response of a 30 nm
thick continuous reference Ni film. The enhancement can be attributed
to the proximity of the used wavelength ($\lambda = 660$ nm) to
the plasmonic resonances at the metal/air interface (see
Fig.~\ref{fig_2}(b)).  TMOKE is featureless when away from the excitation
region. We can understand this behavior if we consider that TMOKE
can be approximated as the product of two
terms:\cite{PhysRevB.76.153402} one is the frequency derivative of
the reflection (transmission) spectrum and the other is the
magnetic field induced frequency shift.  Hence, TMOKE is expected to
be enhanced near the SPPs resonant wavelength even for the case of
smooth interface as it has been observed  in continuous thin films
of Ni in early works.\cite{Ferguson197791}

\begin{figure}
 \includegraphics[width=0.9\columnwidth]{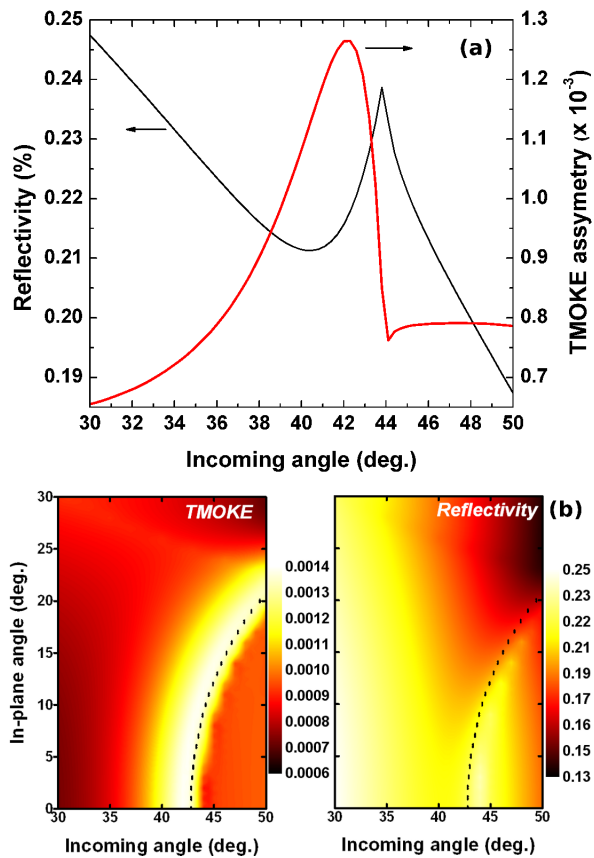}
\caption{\label{fig_4} (Color online) (a) Calculated reflectivity
in the angular range of the the minimum in the $\varGamma K$
direction (black line, left axis) and the TMOKE (red line, right
axis). (b)  Calculated reflectivity (right panel) and TMOKE (left
panel) in the same angular range as in (a) along different
in-plane directions, going from the $\varGamma K$ (0 degrees) to
the  $\varGamma M$ (30 degrees) along the $KM$ direction.}
\end{figure}

To get further insight we have performed numerical simulations of
the reflectivity and of the TMOKE signal, using a recently
developed formalism that uses the scattering matrix approach
adapted to deal with arbitrary orientations of the
magnetization,\cite{PhysRevB.85.245103} using the same elements for
the dielectric tensor as for the Polar Kerr
effect. The simulated results depicted in Fig.~\ref{fig_4}(a) are
in very good agreement with the experimental findings. Moreover, in
Fig.~\ref{fig_4}(b) we show the evolution of the reflectivity
minimum and the TMOKE maximum as a function of the in-plane angle.
As one can see, the position of these two features follows nicely
the dispersion relation of the relevant Bragg plasmon, which is
shown as a dashed line in both panels. This clearly shows that the
plasmon excitation is responsible for these two features. Furthermore,
Fig.~\ref{fig_4}(b) shows that outside the excitation region the
reflectivity and TMOKE signals are featureless and the system behaves
as a uniform medium made of a mixture of Ni and air.

In summary, we have shown that a Ni thin film patterned in a 2D hexagonal
antidot array supports the excitation of SPPs both at the air/Ni
interface and Ni/substrate interface. Transmission and reflection
measurements reveal the excitation of SPPs with broader resonances
due to the high damping of SPPs in ferromagnets. The dispersion
relation of SPPs defined by the hexagonal pattern can be modified
with an application of a transverse magnetic field as the TMOKE
modulated reflectivity signal reveals. The enhancement and the
shift of TMOKE curve is significant and can be altered by the
incoming wavelength and angle of incidence. These effects
pave the road for the development of new optical components and
sensors with great application potential.

The authors would like to thank Piotr Patoka for the preparation
of the polystyrene bead template. The authors acknowledge the
support of the Swedish Research Council (VR), the Knut and Alice
Wallenberg Foundation (KAW) and the Swedish Foundation for
International Cooperation in Research and Higher Education
(STINT). BC and AG-M acknowledge funding from the EU
(NMP3-SL-2008-214107-Nanomagma), the Spanish MICINN (``MAPS''
MAT2011-29194-C02-01, and ``FUNCOAT'' CONSOLIDER INGENIO 2010
CSD2008-00023), and the Comunidad de Madrid (``MICROSERES-CM''
S2009/TIC-1476). JCC acknowledges financial support from the
Spanish MICINN (Contract No.\ FIS2011-28851-C02-01).


%

\end{document}